\documentclass[
prd
,showpacs ,amssymb,nobibnotes,aps,eqsecnum]{revtex4}
\usepackage{graphicx}
\input{epsf}

\usepackage{amsmath,amssymb}
\usepackage{bm}

\newcommand{\dalm}{\kern1pt\vbox{\hrule height 0.9pt\hbox{\vrule width 0.9pt
\hskip 2.5pt\vbox{\vskip 5.5pt}\hskip 3pt\vrule width 0.3pt}\hrule height 0.3pt}
\kern1pt}

\newcommand{\gsim}{\, \raisebox{-0.8ex}{$\stackrel{\textstyle >}{\sim}$ }}

\begin{document}



\title{Signatures of hadron-quark mixed phase in gravitational waves }

\author{Hajime Sotani$^1$} \email{hajime.sotani@nao.ac.jp}
\author{Nobutoshi Yasutake$^1$} 
\author{Toshiki Maruyama$^2$}
\author{Toshitaka Tatsumi$^3$}
\affiliation{
$^1$Division of Theoretical Astronomy, National Astronomical Observatory of Japan, 
2-21-1 Osawa, Mitaka, Tokyo 181-8588, Japan\\
$^2$Advanced Science Research Center, Japan Atomic Energy Agency, Tokai, Ibaraki 319-1195, Japan\\
$^3$Department of Physics, Kyoto University, Kyoto 606-8502, Japan
}

\date{\today}

\begin{abstract}
We calculate stellar oscillations including the hadron-quark mixed phase considering the finite size effects. We find that it is possible to distinguish whether the density discontinuity exists or not in the stars, even if one will observe the gravitational waves of the fundamental mode. 
Additionally, the normalized eigenfrequencies of pressure modes depend strongly on the stellar mass and on the adopted equation of state. Especially, in spite of the fact that the radius of the neutron star with $1.4M_\odot$, which is standard mass, is almost independent from the equation of state with quark matter, the frequencies of pressure modes depend on the adopted equation of state. Thus, via observing the many kinds of gravitational waves, it will be possible to make a restriction on the equation of state.
\end{abstract}

%
\maketitle
\section{Introduction}
\label{sec:I}

In order to detect the gravitational waves, which are oscillations of spacetime itself, several ground-based detectors, such as LIGO, VIRGO, TAMA300, and GEO600, are in operation and there are some projects to build the next generation of detector \cite{Barish2005}. In addition to the ground-based detectors, it is also considering to launch detectors in space like LISA \cite{LISA} and DECIGO \cite{DECIGO}. Since the permeability of gravitational waves could be extremely strong, one can expect to see raw information of the wave sources via gravitational waves. On the other hand, the most promising sources of gravitational waves might be supernovae and mergers of binary compact objects, i.e., the gravitational waves are related to the compact stars, around which the gravitational field should be strong. So, the direct detection of gravitational waves could enable us not only to collect the astronomical data and to reveal the true properties of dense matter \cite{AK1996,AK1998,Kokkotas2001,Sotani2001,MPBGF2003,Sotani2003,Miltos2008,Erich2009}, but also to prove the gravitational theory in the strong-field regime \cite{Sotani2004,Sotani2009a}.

In fact, an attempt to estimate the stellar parameters, such as mass, radius, and equation of state (EOS), via their oscillation properties is not a brand-new idea. In astronomy, the helioseismology has been already established and one could see the interior information of our sun through its oscillation properties. Since the late 1990s, it has been suggested possible to reveal the compact star properties with observing the oscillation spectra,  which is called gravitational wave asteroseismology \cite{AK1996,AK1998}. Furthermore, the detailed analysis of the emitted gravitational waves might permit us to determine the radius of accretion disk around supermassive black hole \cite{Sotani2006} or to see the magnetic effect during the stellar collapse \cite{SYK2007}.

With respect to the neutron stars, the density in the vicinity of the stellar center could become much more than the standard nuclear density, which is around $\rho_0\approx 0.17$ fm$^{-3}$. Since such high density cannot be realized on the earth, the detailed matter properties in the neutron stars are still unknown. However, this means conversely that the neutron stars can be good candidates to know the matter properties in the extreme high density region, where it is suggested that the exotic components of matter, such as hyperons, meson condensates, and quark matter, could appear \cite{HH2000}. The existence of these exotic matter changes the EOS and neutron star structures dramatically \cite{burgio2002, TSHT2007,Yasutake2009}. Namely, as mentioned the above, observing the stellar oscillations and/or the corresponding gravitational waves will tell us the information about the matter properties of neutron stars. Especially, it might be impossible to probe the true properties of matter deep inside the star by any other experiments.

The hyperons are considered to appear at around $2-3\rho_0$, if the nuclear matter would be in beta-equilibrium \cite{schulze2006, ishizuka2008}. On the other hand, there are still many uncertainties with respect to the hadron-quark phase transition, e.g., the EOS of quark matter or a deconfinement mechanism. The presence of quark matter inside the compact objects might play an important role in the astronomical phenomena, for example the backbending effect from phase transition \cite{GPW1997,ZBHG2006}, connection to gamma-ray bursts \cite{DP2007}, the gravitational wave bursts \cite{DPB2006}, the gravitational radiation \cite{OUPT2004,LGG2005,LCCS2006,YKHY2007}, the energy release during the collapse from neutron stars to quark stars \cite{YHE2005, yasutake2010}, the neutrino luminosities~\cite{hatsuda1987, nakazato2008, sagert2009}, and cooling with quark matter \cite{PPLS2000,GBV2005,AJKKR2005,kang2007,SWM2009}. Although the simplest model with the quark matter would be the Maxwell construction, around critical density to appear the quark matter the mixed phase could exist, where baryon number and electric charge should be conserved. Generally, the properties of the mixed phase depend strongly on the electromagnetic interaction and the surface tension, which effects are called ``{\it the finite-size effects}". Due to the finite-size effects, the mixed phase is composed of the nonuniform pasta structure \cite{TSHT2007,Yasutake2009}.  However, it has not been clear how to distinguish the finite size effects by observing the astronomical phenomena. 

In this article, we study the gravitational waves emitted from compact stars with the hadron-quark mixed phase considering the finite size effects. Up to now, there exist a few studies from the gravitational wave asteroseismological point of view, which focus on probing the density discontinuity in the high density region with using specific frequencies of gravitational waves (e.g., \cite{Sotani2001,MPBGF2003}). 
However, no one investigates the effects of the mixed phase between the hadron and quark matter on the specific frequencies. 
So, in this article, preparing the neutron star models with the mixed phase or with density discontinuity between the hadron and quark matter phases,  we will calculate the eigenfrequencies associated with the gravitational waves, where as a first step we adopt the Cowling approximation, i.e., the metric perturbation will be neglected. Then, as varying the stellar properties systematically, we will see the dependence of the existence of quark matter on the oscillation frequencies considering finite size effects. 

This article is organized as follows. In the next section, we describe the equation system to construct the neutron star models and the adopted EOS in this article, and show some stellar models concretely. In section \ref{sec:III}, we derive the perturbation equations with Cowling approximation. With appropriate boundary conditions, the problem to solve becomes the eigenvalue problem.  Then the obtained oscillation spectra will be shown in section \ref{sec:IV}. At last, we make a conclusion in section \ref{sec:V}. In this article, we adopt the unit of $c=G=1$, where $c$ and $G$ denote the speed of light and the gravitational constant, respectively, and the metric signature is $(-,+,+,+)$.

\section{Neutron Star Models}
\label{sec:II}

The equilibrium configurations of non-rotating relativistic stars are spherically symmetric solutions of the well-known Tolman-Oppenheimer-Volkoff (TOV) equations. The metric can be expressed as
\begin{equation}
 ds^2 = -e^{2\Phi} dt^2 + e^{2\Lambda} dr^2 + r^2 d\theta^2 + r^2 \sin^2\theta d\phi^2,
\end{equation}
where $\Phi$ and $\Lambda$ are metric functions with respect to $r$. A mass function $m(r)$ are defined as $m(r)=r(1-e^{-2\Lambda})/2$, which satisfies 
\begin{equation}
  \frac{dm}{dr} = 4\pi r^2\rho,
\end{equation}
where $\rho$ is the energy density, while the TOV equations to determine the distributions of the pressure $P(r)$ and metric function $\Phi(r)$ are described as
\begin{align}
  \frac{dP}{dr} &= -(\rho+P)\frac{d\Phi}{dr}, \\
  \frac{d\Phi}{dr} &= \frac{m+4\pi r^3 P}{r(r-2m)}.
\end{align}
To close the equation system, one needs an additional equation, i.e., the equation of state (EOS).

In this article, we adopt the EOS with the hadron-quark mixed phase with hyperons considering finite size effects according to \cite{TSHT2007,Yasutake2009}. Our EOS for hadron is in the framework of the nonrelativistic Brueckner-Hartree-Fock approach including hyperons such as
$\Sigma^-$ and $\Lambda$~\cite{Baldo1998}. It is not clear now whether the $\Sigma^-$-$N$ interaction is repulsive or not~\cite{noumi02,saha04}, while we use here a weak but attractive interaction. It would be interesting to see how our results are changed by using the other $\Sigma^-$-$N$ interactions, and we will discuss it in the future work. For the comparisons,  we also adopt the EOS composed of only nucleon. We called them as ``hyperon EOS" and ``nucleon EOS" in this article.

For the quark phase, we adopt the MIT bag model. It should be noticed that the adopted EOS in this article is not simple MIT bag model but more sophisticated models suggested in the previous articles \cite{chen09, yasutake09a}. Assuming massless $u$ and $d$ quarks and $s$ quarks with the current mass of $m_s=150$ MeV, we set the bag constant $B$ to be 100~MeV fm$^{-3}$.

For the mixed phase, we assume non-uniform structures, so-called ``pastas". In practice, the structures such as droplet, rod, slab, tube, and bubble are considered. We use the local density approximation for particles using the Wigner-Seitz cell. In order to construct such pasta phase, we put a sharp boundary with a constant surface tension parameter between the hadron and quark phases. Then the Gibbs conditions are imposed, where one needs to solve the chemical equilibrium among particles in two phases consistent with the Coulomb potential, and a pressure balance consistent with the surface tension in self-consistently. Although the knowledge of the value of the surface tension at the hadron-quark interface is very poor, some theoretical estimations have done and they suggest that the value of surface tension is around $\sigma\approx 10-100$ MeV/fm$^2$ \cite{Farhi1984,Kajantie1991}. Since one can see that the models with $\sigma\gsim 40$ MeV/fm$^2$ are almost same as that with $\sigma=40$ MeV/fm$^2$, in this article we consider only a range of $10\le\sigma\le 40$ MeV/fm$^2$. As the extreme case, we also consider the EOS with the Maxwell construction, which has the sharp discontinuity of the density between $5.93\times 10^{14}$ and $8.82\times 10^{14}$  g/cm$^3$. Note that this discontinuity appears at strong surface tensions considering the finite size effects, e.g., $\sigma > 70$ MeV/fm$^2$ in our previous study~\cite{TSHT2007}. Moreover, we take into account another extreme case of the EOS with the bulk Gibbs condition, which appears at the zero surface tension limit \cite{TSHT2007}.  Finally, in order to determine which structure is most favored in the mixed phase, we compare the energy among the pasta structures. It notes that we do not take into account the anti-particles and muons in this article, because their effects should not be so important.

\begin{figure}[htbp]
\begin{center}
\includegraphics[scale=0.45]{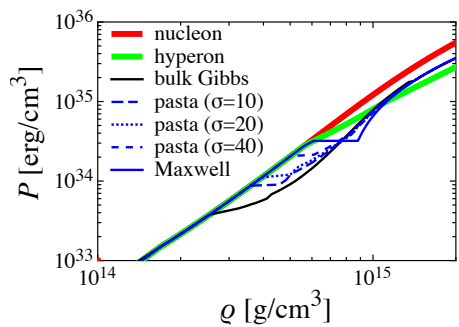} 
\end{center}
\caption{
Relationship between the total energy density including masses~($\rho$) and the pressure~($P$) for the EOSs adopted in this article.
}
\label{fig:EOS}
\end{figure}

\begin{figure}[htbp]
\begin{center}
\includegraphics[scale=0.45]{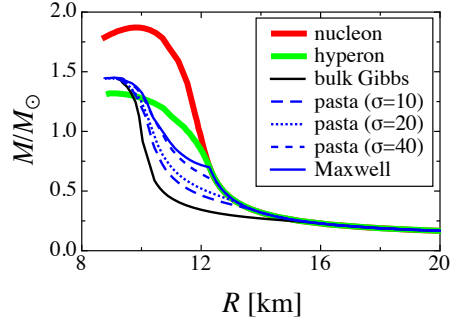} 
\end{center}
\caption{
Mass-radius relation for neutron stars with several EOSs.
}
\label{fig:MR}
\end{figure}

Then the above EOSs should be connected to the hadronic EOS proposed by Negele and Vautherin \cite{EOS-NV} when the density become lower than around $10^{14}$ g/cm$^3$. In Table \ref{tab:EOS}, the components of each EOS adopted in this article are summarized, while Fig. \ref{fig:EOS} shows the relationship between the energy density $\rho$ and the pressure $P$ in the higher density region for the adopted EOSs in this article. Solving the TOV equations with such EOSs, we can get the stellar properties as Fig. \ref{fig:MR}. It should be noticed that the maximum masses of the stellar models composed of EOS including the quark matter are almost independent from the value of $\sigma$, which are around $1.4M_\odot$. At the end of this section, we should mention about the maximum mass of neutron stars. Recently two pulsar mass measurements appeared, which are well above $1.4M_\odot$, i.e., $M=1.97M_\odot$ for PSR J1614-2230 \cite{Demorest2010} and $M=1.667M_\odot$ for PSR J1903+0327 \cite{Freire2010}. Since it is impossible to explain these evidences with the EOS including the quark matter adopted in this article, we should derive more realistic EOS and examine the stellar oscillations as a future work.

\begin{table}[htbp]
\begin{center}
\leavevmode
\caption{Components of each EOS adopted in this article and the references.
}
\begin{tabular}{ccc | ccc }
\hline\hline
  & EOS & & & components \\
\hline
  & nucleon &  &  & nucleon \\
  & hyperon &  &  & nucleon, hyperon  \\
  & bulk Gibbs &  &  & nucleon, hyperon, quark \\
  & with pasta phase &  &  & nucleon, hyperon, quark\\
  & Maxwell &  &  & nucleon, hyperon, quark \\
\hline\hline
\end{tabular}
\label{tab:EOS}
\end{center}
\end{table}

%

%

\section{Perturbation Equations}
\label{sec:III}

In order to determine the specific oscillation frequencies in the neutron stars, in this section we present the perturbation equations for nonradial oscillations of spherically symmetric neutron stars. Especially, in this article we consider only stellar oscillations with Cowling approximation, where the fluid would oscillate on a fixed background metric. Namely, the spacetime will be frozen such that the metric perturbation should be neglected ($\delta g_{\mu\nu}=0$). Thus, with Cowling approximation, our study limits to the modes related to the fluid perturbations, i.e., $f$, $p$, and $g$-modes, while we cannot see the emission of gravitational waves associated with the so-called $w$-modes which correspond to the spacetime oscillations. It should be emphasized that even with this simple approximation one can see qualitatively the features for oscillation frequencies of emitted gravitational waves.

The fluid Lagrangian displacement vector is given by
\begin{equation}
  \xi^i = \left(e^{-\Lambda}W, -V\partial_\theta, -V\sin^{-2}\theta\partial_\phi\right)r^{-2}Y_{\ell m},
\end{equation}
where $W$ and $V$ are functions with respect to $t$ and $r$ while $Y_{\ell m}$ is the spherical harmonic function.
Then the perturbations of the four-velocity, $\delta u^\mu$, can be written as
\begin{equation}
  \delta u^\mu = \left(0, e^{-\Lambda}\partial_t W, -\partial_t V \partial_\theta, -\partial_t V\sin^{-2}\theta\partial_\phi\right)r^{-2}e^{-\Phi}Y_{\ell m}.
\end{equation}
With these variables, the perturbation equations describing the fluid oscillations can be obtained by taking a variation of the energy-momentum conservation law, i.e., $\delta\left(\nabla_\nu T^{\mu\nu}\right)=0$. This equation reduces to $\nabla_\nu\delta T^{\mu\nu}=0$ with Cowling approximation. 
The explicit forms with $\mu=r,\theta$ are
\begin{align}
  &\frac{\rho+P}{r^2}e^{\Lambda-2\Phi}\ddot{W} - \partial_r\left[\frac{\gamma P}{r^2}\left\{e^{-\Lambda}W' + \ell(\ell+1)V\right\}+e^{-\Lambda}P'\frac{W}{r^2}\right]
       \nonumber \\
       &\hspace{2cm} + \frac{P'}{r^2}\left(1+\frac{dP}{d\rho}\right)\left[e^{-\Lambda}W'+\ell(\ell+1)V\right] - \frac{\rho'+P'}{r^2}\Phi'e^{-\Lambda}W = 0, \label{Tr} \\
  &(\rho + P)e^{-2\Phi}\ddot{V} + \frac{\gamma P}{r^2}\left[e^{-\Lambda}W' + \ell(\ell+1)V\right] + \frac{P'}{r^2}e^{-\Lambda}W = 0, \label{Ttheta}
\end{align}
where primes and dots on the variables denote the partial derivative with respect to $r$ and $t$, respectively.  $\gamma$ is the adiabatic constant defined as
\begin{equation}
  \gamma \equiv \left(\frac{\partial\ln P}{\partial\ln n}\right)_s = \frac{n\Delta P}{P\Delta n}, \label{gamma}
\end{equation}
where $n$ is the baryon number density and $\Delta$ denotes the Lagrangian variation. The Lagrangian variation of the baryon number density, $\Delta n$, is determined by the relation as $\Delta n/n=-\nabla_k^{(3)}\xi^k-\delta g/g$, where $\nabla_k^{(3)}$ and $g$ are the covariant derivative in a three dimension with metric $g_{\mu\nu}$ and the determinant of $g_{\mu\nu}$, respectively. Since in this article we adopt the Cowling approximation, the second term is neglected.
Then the Lagrangian variation of $n$ can be described as
\begin{equation}
  \frac{\Delta n}{n} = -e^{-\Lambda}\frac{W'}{r^2} -\frac{\ell(\ell+1)}{r^2}V. \label{dn}
\end{equation}

Assuming a harmonic dependence on time, the perturbative variables will be written as $W(t,r)=W(r)e^{i\omega t}$ and $V(t,r) = V(r)e^{i\omega t}$. Additionally, calculating the combination of the form $d[\rm{Eq.} (\ref{Ttheta})]/dr-[\rm{Eq.} (\ref{Tr})]$ and substituting Eq. (\ref{Ttheta}) again, one can get the simple equation as
\begin{equation}
  V' = 2\Phi'V-e^\Lambda\frac{W}{r^2}. \label{V'}
\end{equation}
Thus, from Eqs. (\ref{Ttheta}) and (\ref{V'}), one can obtain the following simple equation system for the fluid perturbations;
\begin{align}
  W' &= \frac{d\rho}{dP}\left[\omega^2 r^2e^{\Lambda-2\Phi}V + \Phi' W\right] - \ell(\ell+1)e^{\Lambda}V, \label{eq1} \\
  V' &= 2\Phi'V-e^\Lambda\frac{W}{r^2}.  \label{eq2}
\end{align}

In order to solve this equation system, we have to impose appropriate boundary conditions at the stellar center ($r=0$) and at the stellar surface ($r=R$). With these boundary conditions, the problem to solve becomes an eigenvalue problem for the parameter $\omega$. From the above equation system, one can find the behavior of $W$ and $V$ in the vicinity of stellar center as $W(r)=Cr^{\ell+1}+{\cal O}(r^{\ell+3})$ and $V(r) = -Cr^\ell/\ell+{\cal O}(r^{\ell+2})$, where $C$ is an arbitrary constant. On the other hand, the boundary condition at the stellar surface is the vanishing of the Lagrangian perturbation of pressure, i.e., $\Delta P=0$. Since $\Delta P$ could be expressed that $\Delta P=\gamma P \Delta n/n$ from Eq. (\ref{gamma}), with the help of Eqs. (\ref{dn}) and (\ref{eq1}), the condition of $\Delta P=0$ becomes as
\begin{equation}
  \omega^2 r^2e^{\Lambda-2\Phi}V + \Phi' W = 0.
\end{equation}
Furthermore, if one would consider the stellar models with density discontinuity, one has to prepare the additional junction condition at the surface of discontinuity, which are the continuous condition for $W$ and $\Delta P$ \cite{Sotani2001}. These junction conditions can be rewritten with the variables $W$ and $V$ as
\begin{align}
  W_+ &= W_-, \\
  V_+ &= \frac{e^{2\Phi}}{\omega^2 {R_g}^2}\left\{\frac{\rho_-+P}{\rho_++P}
       \left[\omega^2{R_g}^2e^{-2\Phi}V_-+e^{-\Lambda}\Phi' W_-\right]-e^{-\Lambda}\Phi' W_+\right\},
\end{align}
where $R_g$ denotes the position of the density discontinuity, and $W_-$, $V_-$, and $\rho_-$ are the vales of $W$, $V$, and $\rho$ at $r=R_g-0$ while $W_+$, $V_+$, and $\rho_+$ are the values of $W$, $V$, and $\rho$ at $r=R_g+0$, respectively.

\section{Oscillation Spectra}
\label{sec:IV}

In this section we examine the stellar oscillations on the stellar models shown in Sec. \ref{sec:II}. Especially, we focus on the stellar models whose mass is in the range of $0.5M_\odot\le M\le M_{\rm max}$ and at $0.1M_\odot$ intervals, where $M_{\rm max}$ is maximum mass for each EOS. Namely, the masses of the stellar models we adopt in this article are $0.5\le M/M_\odot\le1.3$ for hyperon EOS, $0.5\le M/M_\odot \le1.8$ for nucleon EOS, and $0.5\le M/M_\odot\le 1.4$ for the other EOSs. As mentioned the above, the stellar models with Maxwell EOS have the density discontinuity, if the central density is larger than $8.816\times 10^{14}$  g/cm$^3$. That is, for Maxwell EOS, the stellar models with $0.7\le M/M_\odot \le 1.4$ have the density discontinuity, while those with $M/M_\odot =0.5$ and $0.6$ do not have the density discontinuity and such stellar models are same as those with hyperon EOS (see Fig. \ref{fig:MR}).

\begin{figure}[htbp]
\begin{center}
\begin{tabular}{cc}
\includegraphics[scale=0.45]{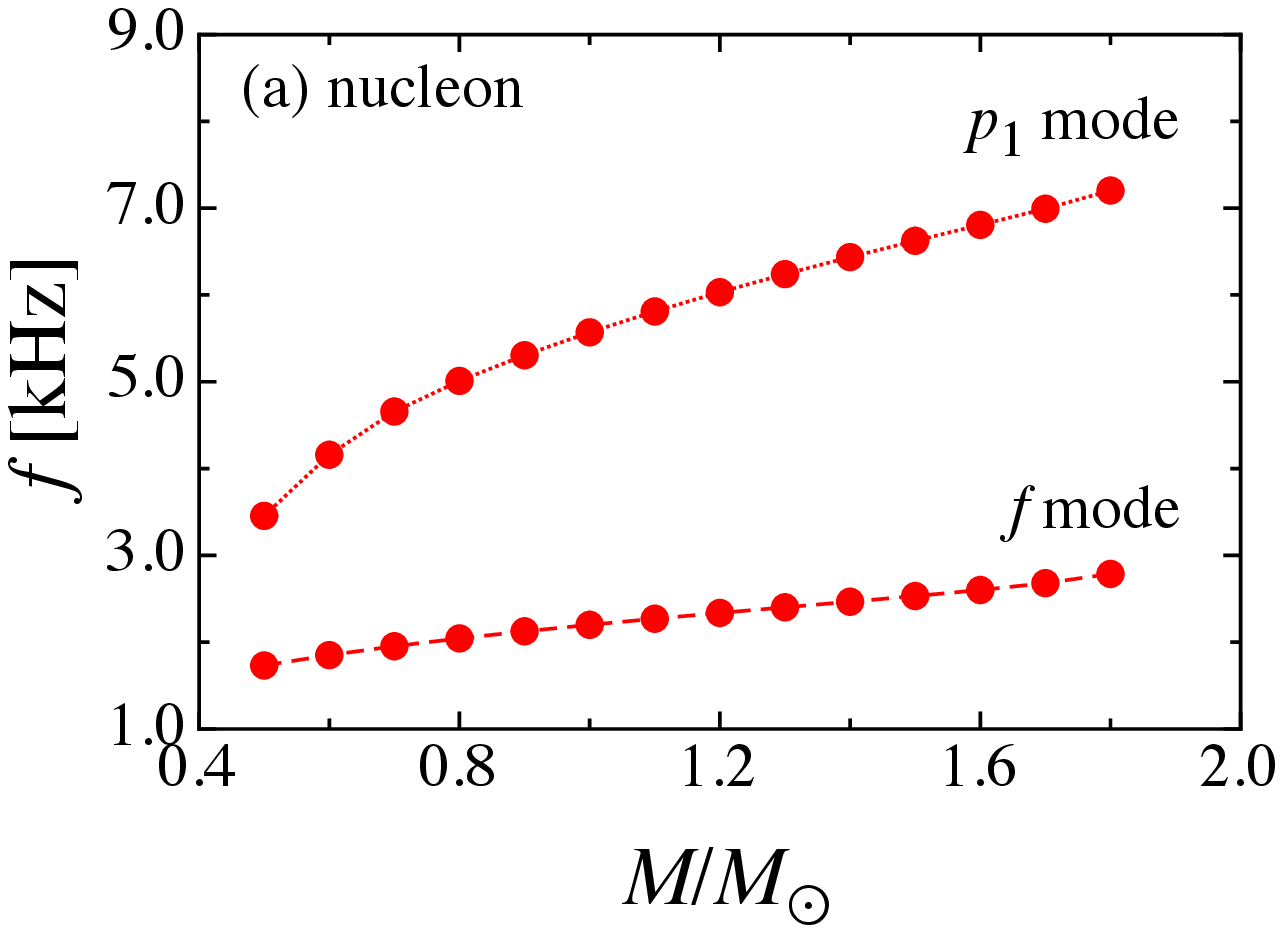} &
\includegraphics[scale=0.45]{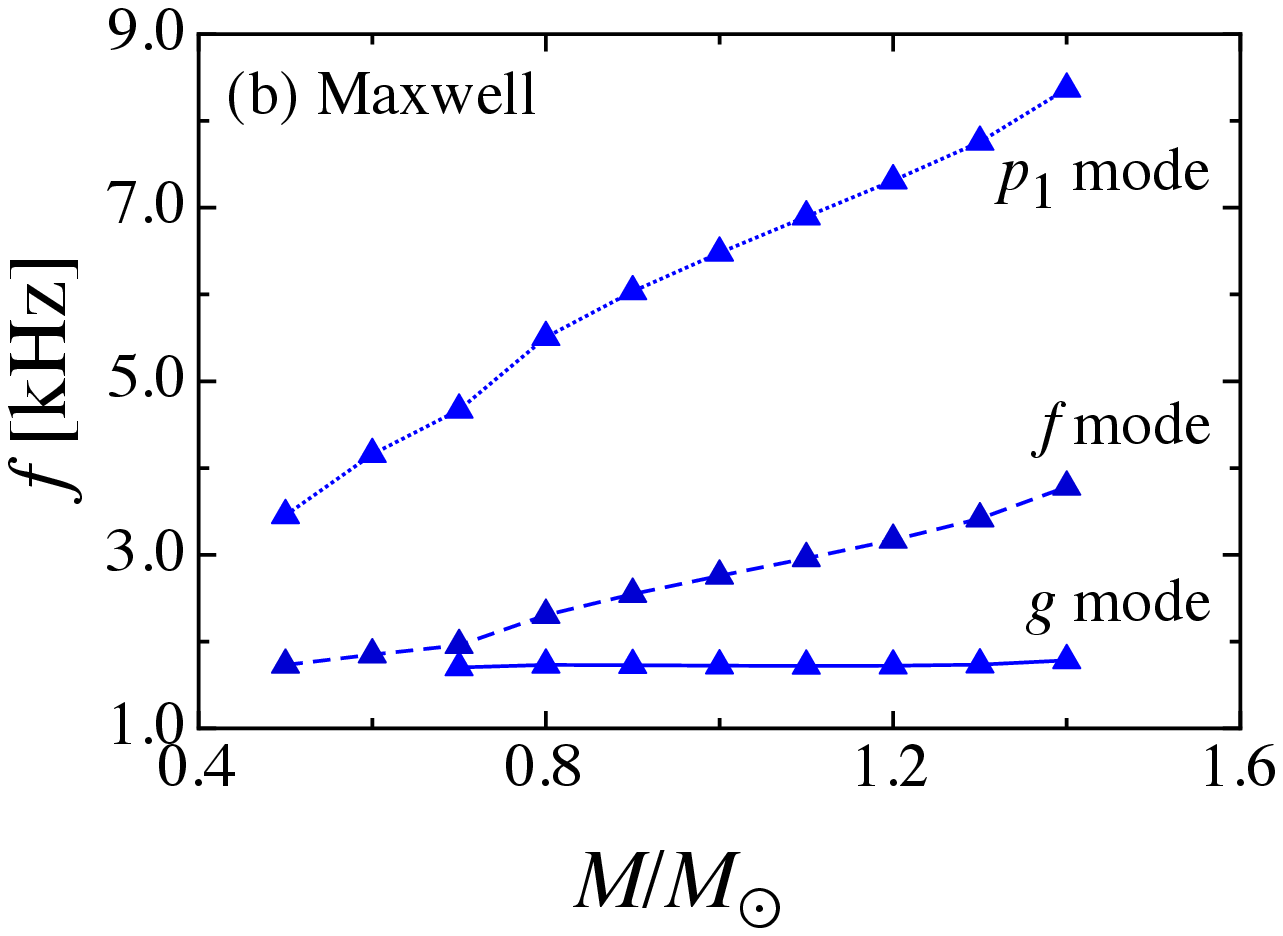} \\
\end{tabular}
\end{center}
\caption{
The first a few eigenfrequencies for the stellar models with (a) nucleon and (b) Maxwell EOSs are plotted as a function of the stellar mass $M/M_\odot$, where the frequency $f$ is defined as $f\equiv \omega/2\pi$. As mentioned in text, for the case of the stellar models with the other EOSs, the kinds of excited eigenfrequencies are same as to the case of the stellar model with nucleon EOS, i.e., they have $f$ and $p_i$ modes, where $i=1,2,3,\cdots$. On the other hand, for the case of the stellar model with Maxwell EOS, which has 1st order phase transition in the density, the additional eigenfrequency are excited, which is $g$ mode.
}
\label{fig:spectrum}
\end{figure}

When neutron stars oscillate, many kinds of gravitational waves are radiated. If the stars are spherically symmetric and without density discontinuity inside the star, which might be the simplest model, the fundamental ($f$), pressure ($p$), and spacetime ($w$) modes are excited, where $f$ and $p$ modes are gravitational waves related to the fluid oscillations while $w$ modes correspond to the oscillations of spacetime itself. If the stars are spherically symmetric and with density discontinuity, the additional oscillation modes, i.e., the $g$ modes, are excited as well as $f$, $p$, and $w$ modes. The $g$ modes are also gravitational waves associated with the fluid oscillations. In this article, we will see qualitatively the gravitational waves related to the fluid oscillations because the Cowling approximation is adopted in our analysis. Thus, as shown in Fig. \ref{fig:spectrum}, the stellar models with the adopted EOS except for Maxwell EOS have $f$ and $p$ mode, while those with Maxwell EOS, whose masses are more than $0.7M_\odot$, have $f$, $p$, and $g$ modes.

%
%

\begin{figure}[htbp]
\begin{center}
\includegraphics[scale=0.45]{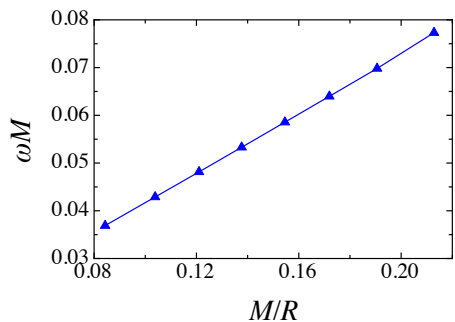} 
\end{center}
\caption{
The normalized eigenvalue $\omega$ of $g$ mode for the stellar models with Maxwell EOS are plotted as a function of the stellar compactness $M/R$.
}
\label{fig:g}
\end{figure}

Before discussing the $f$ and $p$ modes, we pay attention to the $g$ mode for the stellar models with Maxwell EOS. As noted before, $g$ mode is excited due to the existence of density discontinuity. Therefore, one could know the existence of density discontinuity inside the neutron stars, if the $g$ mode gravitational waves will be observed. Actually, since the typical frequency of $g$ mode is in the range from a few hundred Hz up to kHz, such gravitational waves could be observed by using the ground-based gravitational waves detectors. From Fig. \ref{fig:spectrum}, one can observe that the frequency of $g$ mode is almost independent from the stellar mass, which is around $1.73$ Hz. However, we find that $g$ mode frequency can be expressed well as a function of stellar compactness $M/R$ (see Fig. \ref{fig:g}), such as
\begin{equation}
 \omega M = 0.3130 \left(\frac{M}{R}\right)+0.0103.
\end{equation}
Practically, this empirical formula can expect the $g$ mode frequency with less than $0.6$\% accuracy. So, via observing the $g$ mode gravitational waves, one could know the stellar properties. Or, with the observation of redshift parameter, which is connected to the stellar compactness directly, one could make a restriction on the stellar mass.

%
%

\begin{figure}[htbp]
\begin{center}
\begin{tabular}{cc}
\includegraphics[scale=0.45]{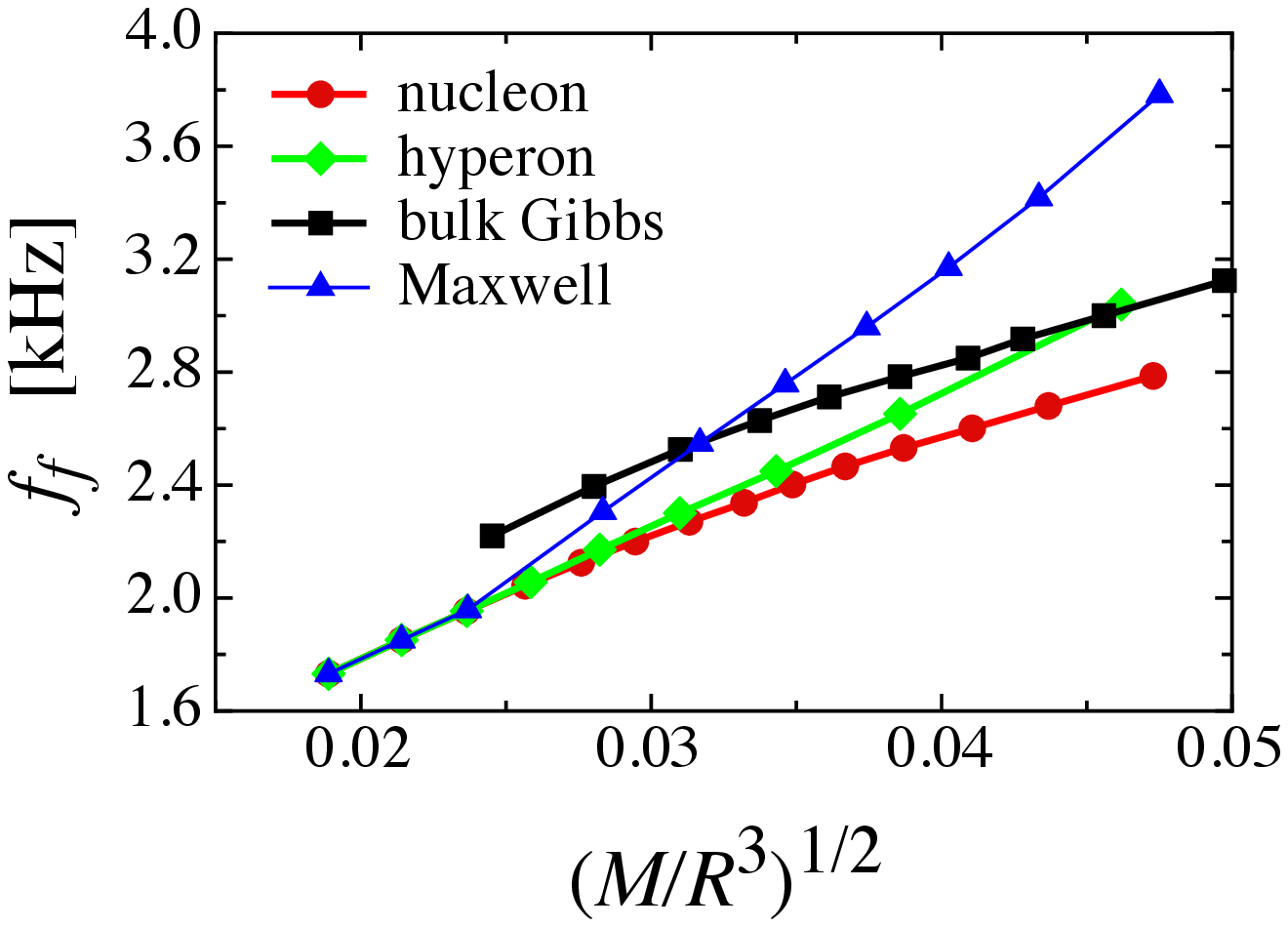} &
\includegraphics[scale=0.45]{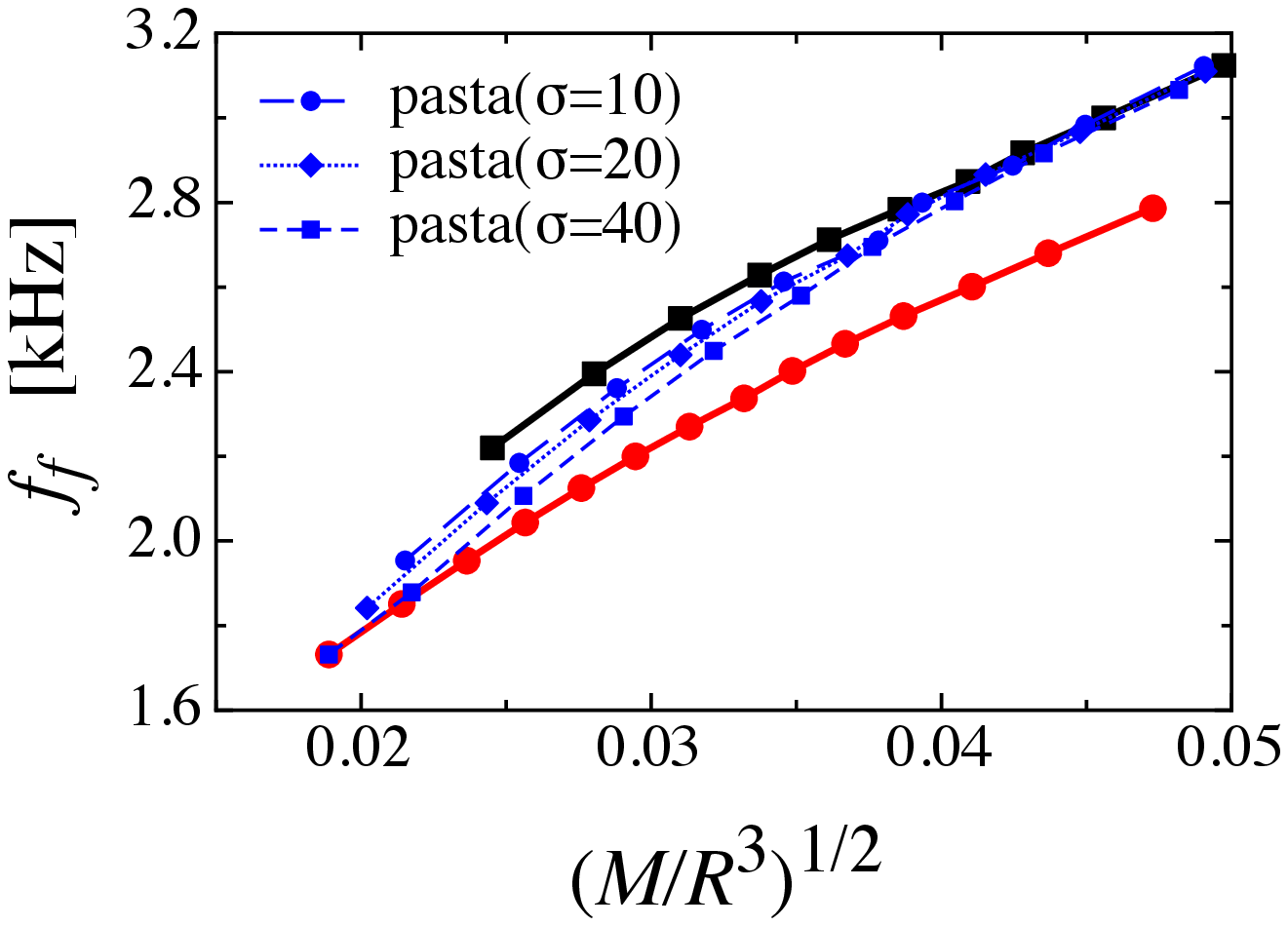} \\
\end{tabular}
\end{center}
\caption{
With several EOSs, the frequencies of $f$ modes are plotted as a function of the stellar average density $(M/R^3)^{1/2}$, where $f_f$ is defined as $f_f\equiv \omega_f/2\pi$. The left panel corresponds to the results with nucleon, hyperon, bulk Gibbs, and Maxwell EOSs, while the right panel focus on the pasta EOSs.
}
\label{fig:f-avdensity}
\end{figure}

On the other hand, it is well-known that the frequency of $f$ mode can be connected to the stellar average density $\left(M/R^3\right)^{1/2}$, which could be physically explained by considering the relation between the sound speed and the propagation time of the fluid perturbation inside the star. In fact, Andersson and Kokkotas derived the empirical formula for the $f$ mode frequency by calculating in the stellar models with the several realistic EOSs, which did not include the EOS considering the quark matter, and found that the obtained $f$ mode frequencies in their article are subject to this empirical formula almost independently of the adopted EOS \cite{AK1998}. While, in Fig. \ref{fig:f-avdensity}, we show the $f$ mode frequencies for the stellar models with adopted EOSs in this article. At a glance, one can observe that the behavior of $f$ mode frequencies for the stellar models with Maxwell EOS is quite different from the others. These deviations of the results for Maxwell EOS become up to $36\%$ from those for nucleon EOS and $27\%$ from those for the other EOSs. This means that one could possible to know the existence of the density discontinuity even by observing the $f$ mode gravitational waves as well as $g$ mode ones. Additionally, Fig. \ref{fig:f-avdensity} shows that the $f$ modes for the stellar models with pasta EOS are similar to those for the stellar models with the other EOS without density discontinuity.

%
%

\begin{figure}[htbp]
\begin{center}
\begin{tabular}{cc}
\includegraphics[scale=0.45]{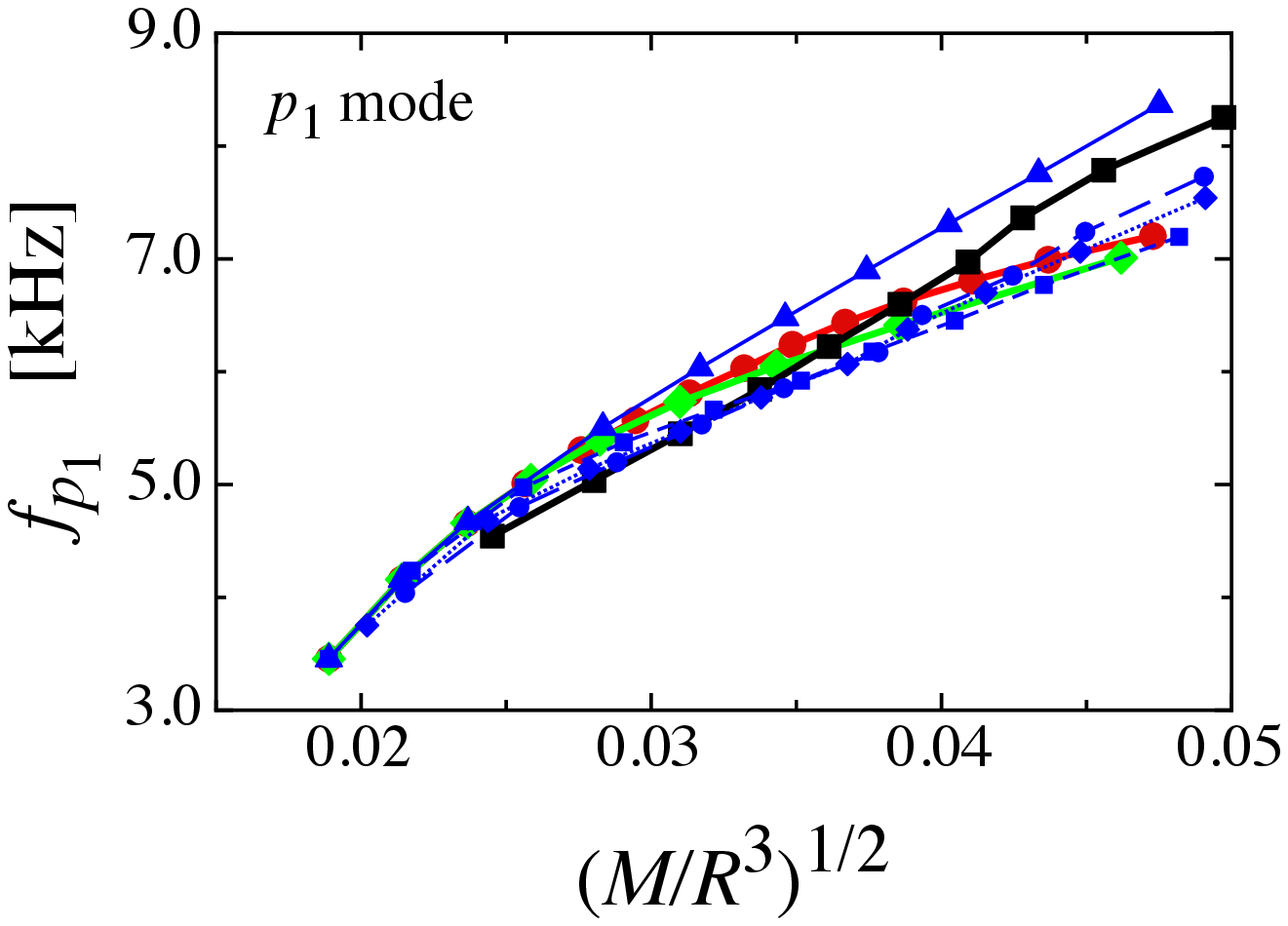} &
\includegraphics[scale=0.45]{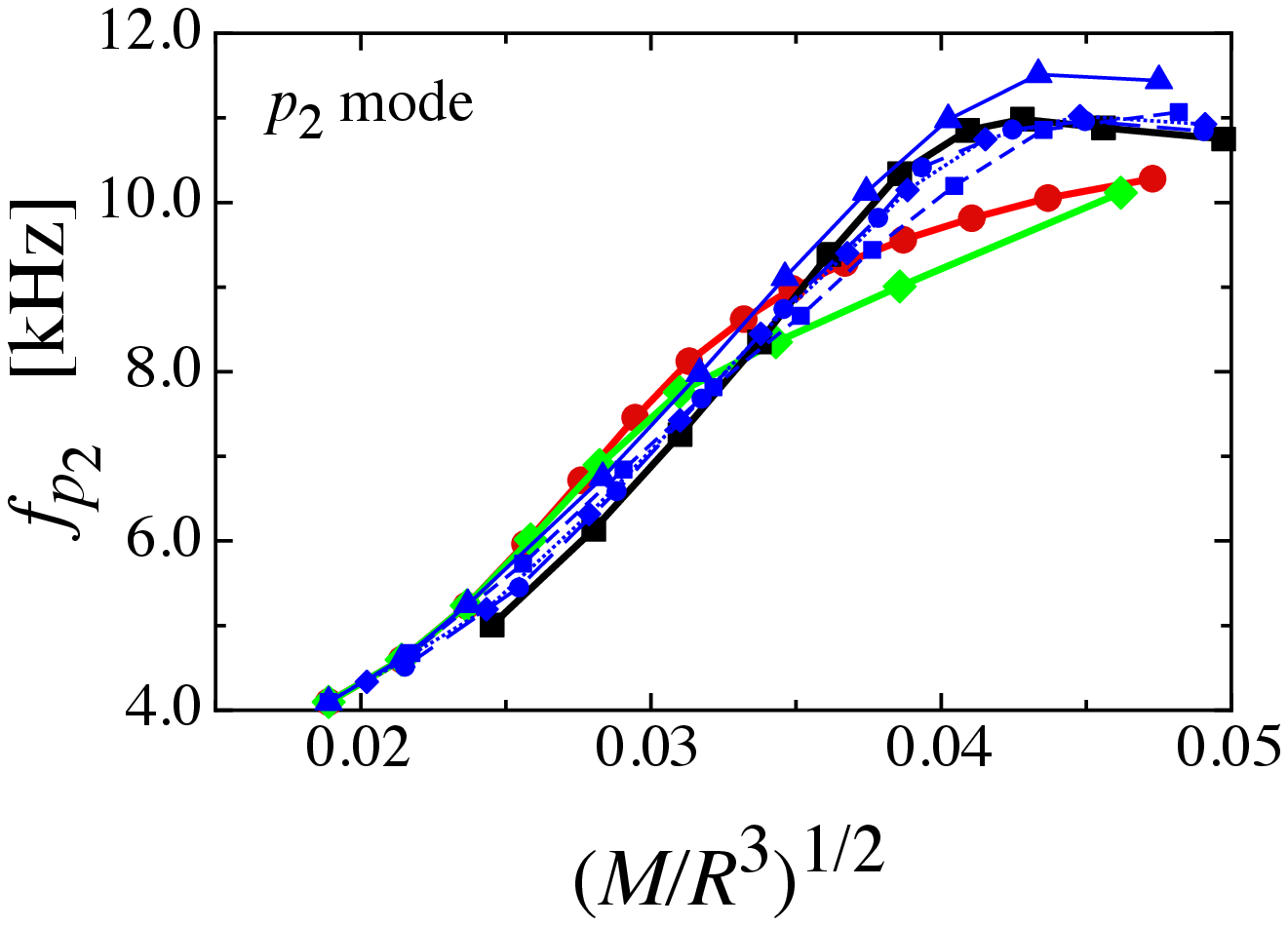} \\
\end{tabular}
\end{center}
\caption{
With several EOSs, the frequencies of $p_1$ (left panel) and $p_2$ (right panel) modes are plotted as a function of the stellar average density $(M/R^3)^{1/2}$, where the marks in figures are corresponding to those in Fig. \ref{fig:f-avdensity}.
}
\label{fig:p-avdensity}
\end{figure}

\begin{figure}[htbp]
\begin{center}
\begin{tabular}{ccc}
\includegraphics[scale=0.42]{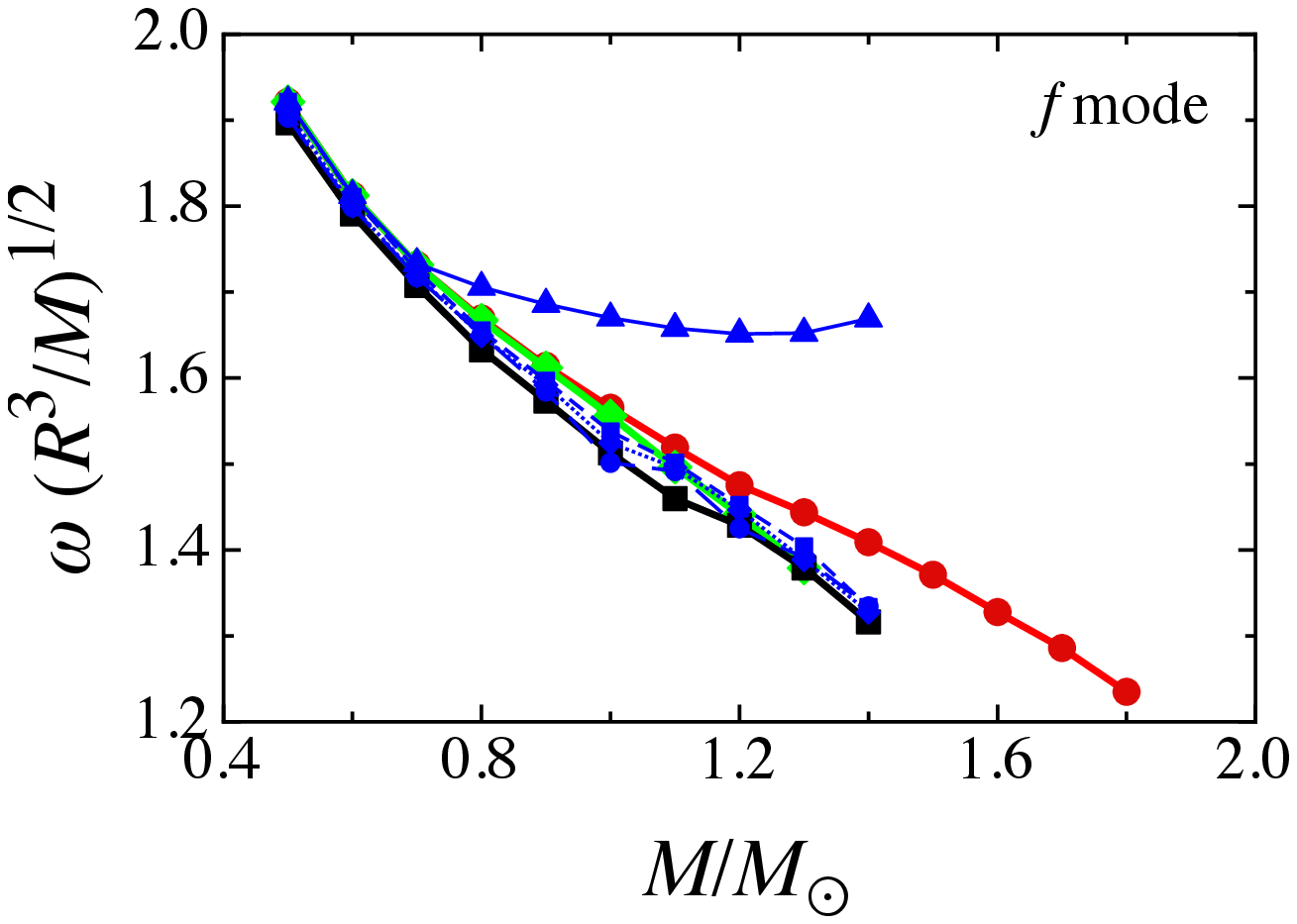} &
\includegraphics[scale=0.42]{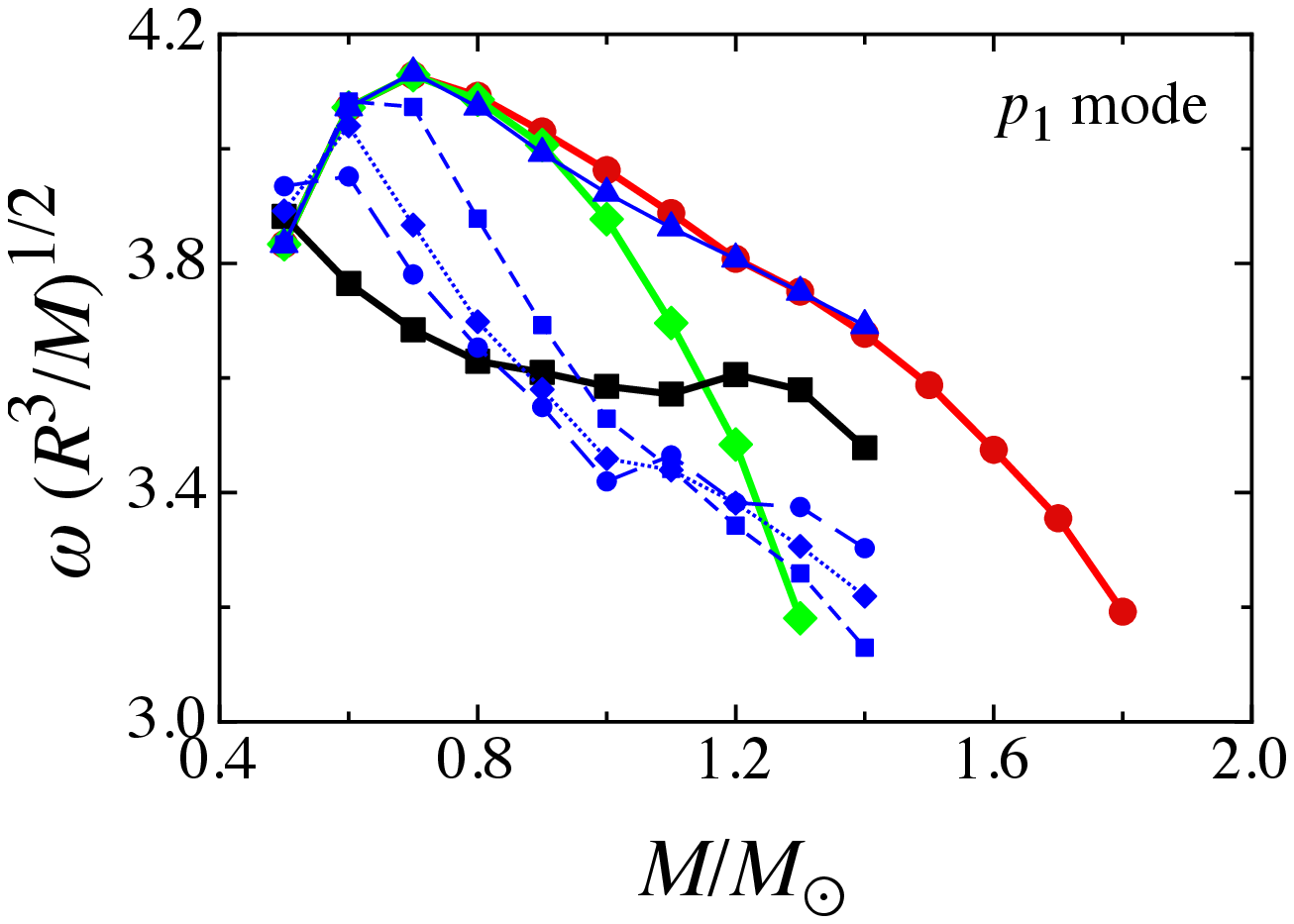} &
\includegraphics[scale=0.42]{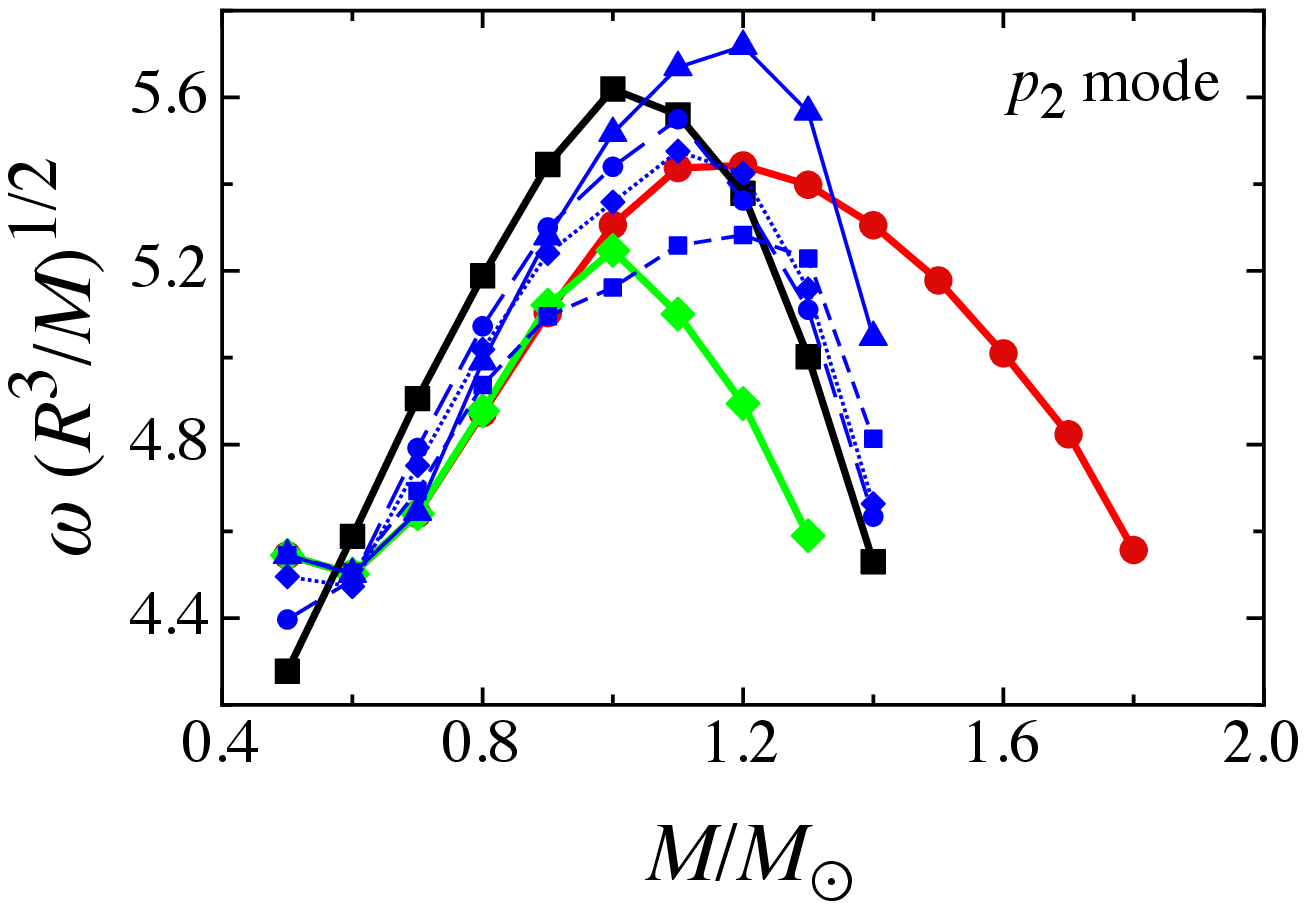} \\
\end{tabular}
\end{center}
\caption{
The normalized eigenvalues of $f$ (left panel), $p_1$ (middle panel), and $p_2$ (right panel) modes are plotted as a function of the stellar mass $M/M_\odot$, where the marks in figures are corresponding to those in Fig. \ref{fig:f-avdensity}.
}
\label{fig:mode}
\end{figure}

Regarding to the $p$ modes, we plot the frequencies of $p_1$ and $p_2$ modes as functions of average density in Fig. \ref{fig:p-avdensity}. From these figures, one can see that $p_i$ modes frequencies are almost independent from the EOS even if that includes the density discontinuity. However, one can see the dependence of $p$ mode frequency on the EOS if we make the figure of the normalized eigenvalues with average density, $\omega\left(R^3/M\right)^{1/2}$, as a function of the stellar mass (see Fig. \ref{fig:mode}). The most interested point in this figure is that the normalized eigenvalues for the stellar models with $1.4M_\odot$ depend strongly on the adopted EOS, in spite of the fact that the stellar shapes are almost independent from the adopted EOSs including quark matter, i.e., $R=9.42-9.67$ km. Moreover, in this figure, the dependence of the normalized eigenvalues of $p_1$ mode on EOS is different from that of $p_2$ modes. Thus, with the help of the observations of stellar mass, it could be possible to distinguish the EOS by observing the several kinds of oscillation modes. At last , it should be also noticed that the normalized eigenvalues of $f$ modes for the stellar model with Maxwell EOS are obviously different from the other stellar models as well as Fig. \ref{fig:f-avdensity}.

%
%
%
%
%
%

\section{Conclusion}
\label{sec:V}

We study how to distinguish the finite effects on the hadron-quark mixed phase by observing the gravitational waves, for which we derive the perturbation equations of neutron stars and obtain their eigenfrequencies.

We find that one could know the existence of density discontinuity inside the star via observing the gravitational waves of not only $g$ mode but also $f$ mode. 
Note that this discontinuity comes from the instability of the mixed phase due to the strong surface tension.
Additionally, it is possible to see the stellar properties by observing the $g$ mode frequency, since such frequency can be expressed well as a function of the stellar compactness. If the EOS do not include the density discontinuity, it might be difficult to distinguish the EOS only by observing the $f$ mode frequencies. However, the normalized eigenfrequencies of $p$ modes depend strongly on the EOS even if the EOS do not including the density discontinuity, although the raw frequencies of $p$ mode is almost independent. Thus, with the help of the observation of stellar properties, it could be possible to make a restriction on the stellar EOS.

In this article, for simplicity, we assume the Cowling approximation, which restricts our examination to only stellar oscillations. This means that we should do a more detailed study including the metric perturbations. Via this type of oscillations, we could obtain the additional information, such as the damping rate of gravitational waves, and combining with results shown in this article would provide more accurate constraints on the stellar properties and/or the stellar EOS. Furthermore the stellar magnetic field might play an important role. For example, the quasi-periodic oscillations have observed in the giant flare and these phenomena are believed to be oscillations of strong magnetized neutron stars \cite{Sotani2007}. Considering the effects of stellar magnetic fields, it might be possible to obtain the further information. 

Additionally, although we focus only on neutron star matter~($T=0$ MeV and $Y_{\nu_e}=0$) in this article, in order to study the proto neutron stars, we should take into account other effects, e.g., the thermal effects on the stellar oscillations~\cite{McDermott}, and  the effects of temperature and/or neutrino trapping on the pasta structures~\cite{Yasutake2009, pagliara09, hempel09}.

\acknowledgments
We are grateful to Y.~Sekiguchi,  S. Chiba, H.-J. Schulze, G.~F.~Burgio and M.~Baldo for their warm hospitality and fruitful discussions, and also to referee for the valuable comments. This work was partially supported by the Grant-in-Aid for the Global COE Program ``The Next Generation of Physics, Spun from Universality	and Emergence,'' from the Ministry of Education, Culture, Sports, Science and Technology (MEXT) of Japan, and the Grant-in-Aid for Scientific Research (C) (20540267, 21105512, 19540313).



\end{document}